\documentclass[a4paper,11pt]{article}
\usepackage{pos}
\usepackage{microtype}
\usepackage{float}

\title{Precise measurement of the neutron $d_2$}
\ShortTitle{Precise measurement of $d_2^n$ as a function of $Q^2$}
\author*[a]{Minho Kim}
\onbehalf{on behalf of the $\boldsymbol{A_1^n/d_2^n}$ Collaboration}
\affiliation[a]{Physics Division, Argonne National Laboratory,\\
  9700 S Cass Ave, Lemont, IL 60439, USA}
\emailAdd{minho.kim@anl.gov}

\abstract{
In inclusive polarized lepton--nucleon deep-inelastic scattering, the twist-3 matrix element $d_2$ is defined as the third moment of a linear combination of the spin structure functions $g_1$ and $g_2$. It is sensitive to quark--gluon correlations in the nucleon and has been interpreted as an average color Lorentz force exerted on the struck quark by the remnant diquark system. The neutron $d_2^n$ has been measured by several experiments, but the observables entering these extractions were obtained over broad ranges in $Q^2$, requiring $g_1$ and $g_2$ to be evolved to a common $Q^2$, and limited statistics have left $d_2^n$ at high $Q^2$ inconclusive. The Jefferson Lab E12-06-121 experiment was designed to determine $d_2^n$ precisely by measuring double-spin asymmetries and unpolarized cross sections with high statistics at two nearly constant values of $Q^2$, $\sim 2.5$ and $\sim 5.0$~GeV$^2$, in scattering of longitudinally polarized electrons from a polarized $^3$He target. As a first step in that analysis, we report preliminary results for $g_2^n$ in the DIS region. At both values of $Q^2$, the data lie below the Wandzura--Wilczek prediction over a limited range at lower $x$ and agree with it at larger $x$. Extending the analysis into the resonance region will allow a precise extraction of $d_2^n$.}

\FullConference{33rd International Workshop on Deep Inelastic Scattering and Related Subjects (DIS2026)\\
4--8 May 2026\\
Bologna, Italy}

\hypersetup{pdftitle={Precise measurement of the neutron d2},
  pdfauthor={Minho Kim, on behalf of the A1n/d2n Collaboration}}

\begin{document}
\maketitle

\section{Introduction}
Since the late 1970s, polarized inclusive lepton--nucleon deep-inelastic scattering (DIS) has played a central role in studies of nucleon spin structure through measurements of two spin structure functions, $g_1(x, Q^2)$ and $g_2(x, Q^2)$. Here, $x$ is the Bjorken scaling variable and $Q^2$ is the negative squared four-momentum transfer carried by the exchanged virtual photon. Experimental studies have focused primarily on $g_1$, which probes how quark and antiquark spins contribute to the nucleon spin, while measurements of $g_2$ remain comparatively sparse. Beyond leading twist, $g_2$ receives contributions from quark--gluon
correlations without power suppression~\cite{JaffeJi}, which makes it a direct probe of such correlations in the nucleon. Isolating these contributions requires separating $g_2$ from its twist-2 part, $g_2^{\mathrm{WW}}$, which is determined entirely by $g_1$ through the Wandzura--Wilczek relation~\cite{WW},
\begin{equation}
  g_2^{\mathrm{WW}}(x,Q^2)
  =-g_1(x,Q^2)+\int_x^1\frac{g_1(y,Q^2)}{y}\,dy.
  \label{eq:ww}
\end{equation}
The difference
\begin{equation}
  \bar{g}_2(x,Q^2)=g_2(x,Q^2)-g_2^{\mathrm{WW}}(x,Q^2)
  \label{eq:g2bar}
\end{equation}
therefore carries the higher-twist contribution, and is accessed by measuring $g_2$ precisely and evaluating $g_2^{\mathrm{WW}}$ from existing $g_1$ data.

A quantity characterizing the higher-twist contribution is $d_2$, defined as
\begin{equation}
\begin{aligned}
  d_2(Q^2)
  &= 3\int_0^1 x^2\,\bar{g}_2(x,Q^2)\,dx \\
  &= \int_0^1 x^2
    \left[2g_1(x,Q^2)+3g_2(x,Q^2)\right]\,dx.
\end{aligned}
\label{eq:d2}
\end{equation}
$d_2$ is related to a twist-3 reduced matrix element of a local quark--gluon operator~\cite{JiUnrau} and can be computed in lattice QCD~\cite{Gockeler,Burger}. It can also be interpreted as the average transverse color Lorentz force experienced by the struck quark from the remnant diquark system during the polarized DIS process~\cite{Burkardt}.

Because $d_2$ is evaluated as an integral over $x$ at fixed $Q^2$, a precise determination requires measurements covering a broad range in $x$ at a common $Q^2$. All previous measurements, however, were based on data spanning wide ranges in $Q^2$~\cite{Anthony,Zheng,Posik}, with the structure functions evolved to a common $Q^2$. The Jefferson Lab (JLab) E12-06-121 experiment was designed to provide a significantly improved determination of $d_2^n$ at intermediate and high $Q^2$ for comparison with lattice QCD, with a kinematic coverage spanning a broad range in $x$ at two nearly fixed values of $Q^2$. This contribution reports preliminary $g_2^n$ results in the DIS region as a step toward the full $d_2^n$ extraction.

\section{The E12-06-121 experiment}
The E12-06-121 experiment was carried out in Hall C in 2020. A longitudinally polarized electron beam was scattered from a polarized $^3$He target, and the scattered electrons were detected in the High Momentum Spectrometer (HMS)~\cite{HMS} and the Super High Momentum Spectrometer (SHMS)~\cite{SHMS} to measure the unpolarized cross section $\sigma_0$ and the longitudinal and transverse double-spin asymmetries, $A_{\parallel}$ and $A_{\perp}$. The double-spin asymmetries were formed from the electron yields recorded with opposite beam helicities,
\begin{equation}
  A_{\parallel(\perp)}
  = \frac{1}{P_b P_t D_{\mathrm{N_2}}}\,
    \frac{N^{\downarrow\Uparrow(\Rightarrow)}-N^{\uparrow\Uparrow(\Rightarrow)}}
         {N^{\downarrow\Uparrow(\Rightarrow)}+N^{\uparrow\Uparrow(\Rightarrow)}},
  \label{eq:dsa}
\end{equation}
where $N$ is the number of detected electrons, $P_b$ and $P_t$ are the beam and target polarizations, respectively, and $D_{\mathrm{N_2}}$ is the dilution factor accounting for the unpolarized nitrogen, which was admixed with the $^3$He as a quenching gas during optical pumping. The single arrows denote the electron spin, parallel ($\uparrow$) or antiparallel ($\downarrow$) to the incident electron momentum, and the double
arrows denote the target spin orientation, either longitudinal ($\Uparrow$, along the beam) or transverse ($\Rightarrow$, perpendicular to the beam in the scattering plane).

\begin{figure}[htbp]
  \centering
  \includegraphics[width=0.6\textwidth]{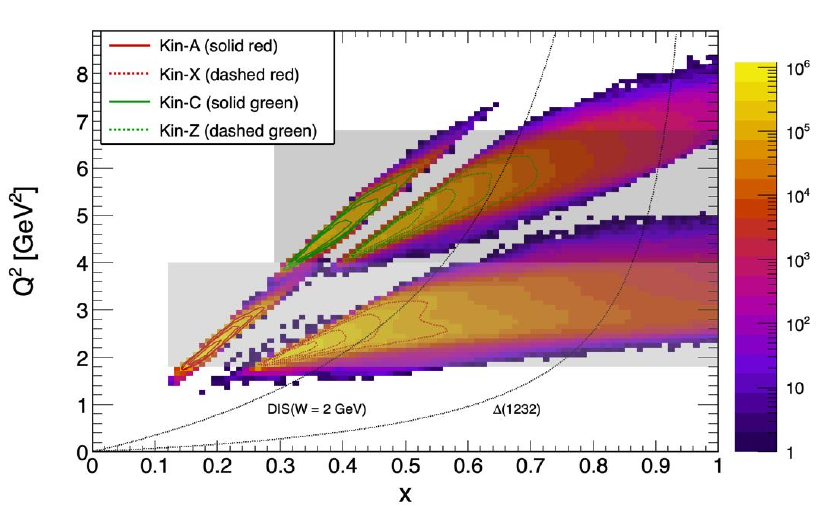}
  \caption{Kinematic coverage of E12-06-121 in the $x$--$Q^2$ plane. Contours indicate the four settings, Kin-A and Kin-X at intermediate $Q^2$ and Kin-C and Kin-Z at high $Q^2$. The dotted curves mark $W=2$~GeV and the $\Delta(1232)$ resonance.}
  \label{fig:xQ2}
\end{figure}

Each spectrometer was operated at two scattering angles in order to cover the intermediate- and high-$Q^2$ regions, $13.5^\circ$ (Kin-A) and $20^\circ$ (Kin-C) for the HMS and $11^\circ$ (Kin-X) and $18^\circ$ (Kin-Z) for the SHMS. The resulting coverage in the $x$--$Q^2$ plane is shown in Fig.~\ref{fig:xQ2}. 

\section{Analysis}
Electrons were selected using the gas Cherenkov detector and electromagnetic calorimeter in each spectrometer. The selection criteria were optimized to retain high electron detection efficiency while suppressing pion backgrounds, with the residual pion contamination estimated to be $10^{-4}$ or less. Another background arises from pair-produced electrons, which was quantified by reversing the spectrometer magnet polarity and detecting positrons under otherwise identical conditions, giving an $e^+/e^-$ ratio of
$10^{-3}$ or below. Two further sources of false asymmetry, the helicity-correlated beam charge asymmetry and the asymmetry in the data-acquisition livetime, were evaluated separately and found to be at or below $10^{-4}$. Events were further required to satisfy $W > 2$~GeV, which selects the DIS region and excludes the resonance region covered at larger $x$.

The measured $\sigma_0$ and $A_{\parallel,\perp}$ were corrected for radiative effects to obtain the Born-level quantities. These effects arise from higher-order QED processes and energy loss as the incident and scattered electrons traverse the experimental materials. The corrections were determined by comparing model calculations of $\sigma_0$ and of the polarized cross section differences $\Delta\sigma_{\parallel,\perp}$ evaluated with and without radiative effects. At the Born level, $\sigma_0$ was taken from the \texttt{F1F221} parameterization~\cite{Christy,Bosted} refitted to world $^3$He cross section data, and $\Delta\sigma_{\parallel,\perp}$ were constructed from the DSSV14 polarized parton distributions~\cite{DSSV}. Both were radiated following the formalism of Mo and Tsai~\cite{MoTsai,Tsai}, and the associated asymmetries were obtained as
\begin{equation}
  A_{\parallel,\perp}=\Delta\sigma_{\parallel,\perp}/(2\sigma_0).
  \label{eq:asym_dsigma}
\end{equation}
The resulting corrections were modest, changing $\sigma_0$ by less than 10\% and shifting $A_\perp$ and $A_\parallel$ by less than $4\times10^{-4}$ and $3\times10^{-3}$ in absolute value.

The corrected $\sigma_0$ and $A_{\parallel,\perp}$ were combined to form the $^3$He spin structure function $g_2^{^3\mathrm{He}}$,
\begin{equation}
  g_2^{^3\mathrm{He}} = \frac{MQ^2}{4\alpha^2}\,
    \frac{y^2\sigma_0}{(1-y)(2-y)}
    \left[-A_{\parallel}
      + \frac{1+(1-y)\cos\theta}{(1-y)\sin\theta}\,A_{\perp}\right],
  \label{eq:g2he3}
\end{equation}
with $M$ the nucleon mass, $\alpha$ the fine-structure constant, $\theta$ the electron scattering angle, and $y=(E-E')/E$ the fraction of the incident electron energy transferred to the target in the laboratory frame, where $E$ and $E'$ are the incident and scattered electron energies. $g_2^n$ was extracted from $g_2^{^3\mathrm{He}}$ based on the effective nucleon polarization approach~\cite{Bissey},
\begin{equation}
  g_2^{n} = \frac{1}{P_n + 0.056}
    \left[g_2^{{}^3\mathrm{He}} - \left(2P_p - 0.014\right)g_2^{p}\right],
  \label{eq:neutron}
\end{equation}
where $P_n$ and $P_p$ are the effective neutron and proton polarizations in $^3$He, respectively, and the constants $0.056$ and $0.014$ account for the $\Delta$-isobar contributions. $g_2^p$ was approximated by $g_2^{\mathrm{WW}}$ evaluated from DSSV14 polarized parton distributions~\cite{DSSV}.

\section{Results}
\begin{figure}[h]
    \centering
    \includegraphics[width=1.0\textwidth]{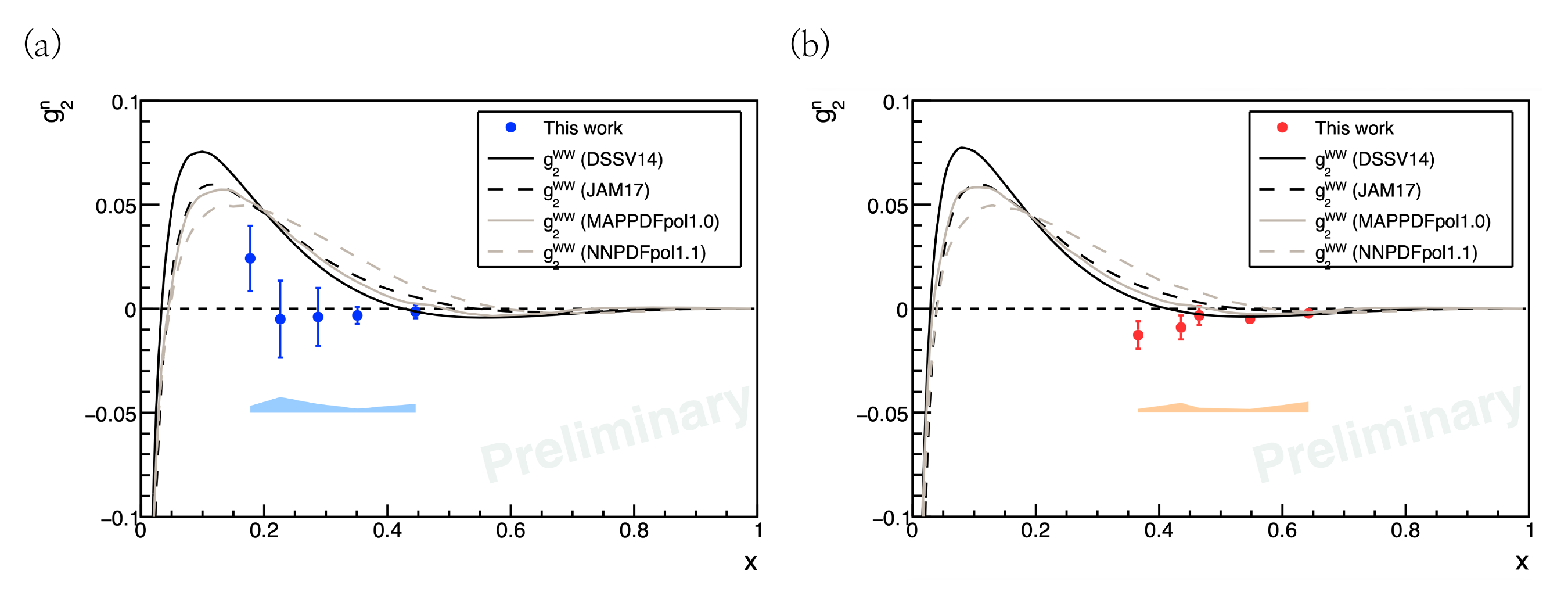}
    \caption{$g_2^n$ as a function of $x$ at (a) $\langle Q^2\rangle$ = 2.5~GeV$^2$ and (b) 5.0~GeV$^2$. Error bars represent statistical uncertainties. The shaded bands below the data points indicate systematic uncertainties. The curves show $g_2^{\mathrm{WW}}$ evaluated from the four global analyses, DSSV14~\cite{DSSV}, JAM17~\cite{JAM}, MAPPDFpol1.0~\cite{MAP}, and NNPDFpol1.1~\cite{NNPDF}.}
    \label{fig:g2n}
\end{figure}

Figure~\ref{fig:g2n} presents $g_2^n$ as a function of $x$ at $\langle Q^2\rangle = 2.5$ and $5.0$~GeV$^2$, together with the $g_2^{\mathrm{WW}}$ predictions from the four global analyses. The systematic uncertainty is dominated by the event selection criteria, comprising the electron identification criteria and the rejection of events originating in the target cell windows. At $\langle Q^2\rangle = 2.5$~GeV$^2$, the data fall below all four $g_2^{\mathrm{WW}}$ predictions over $0.23 \lesssim x \lesssim 0.35$. The data at
$\langle Q^2\rangle = 5.0$~GeV$^2$ behave similarly over $0.37 \lesssim x \lesssim 0.44$. At larger $x$, $g_2^n$ values agree with the predictions within uncertainties at both $Q^2$.

\begin{figure}[h]
    \centering
    \includegraphics[width=0.63\textwidth]{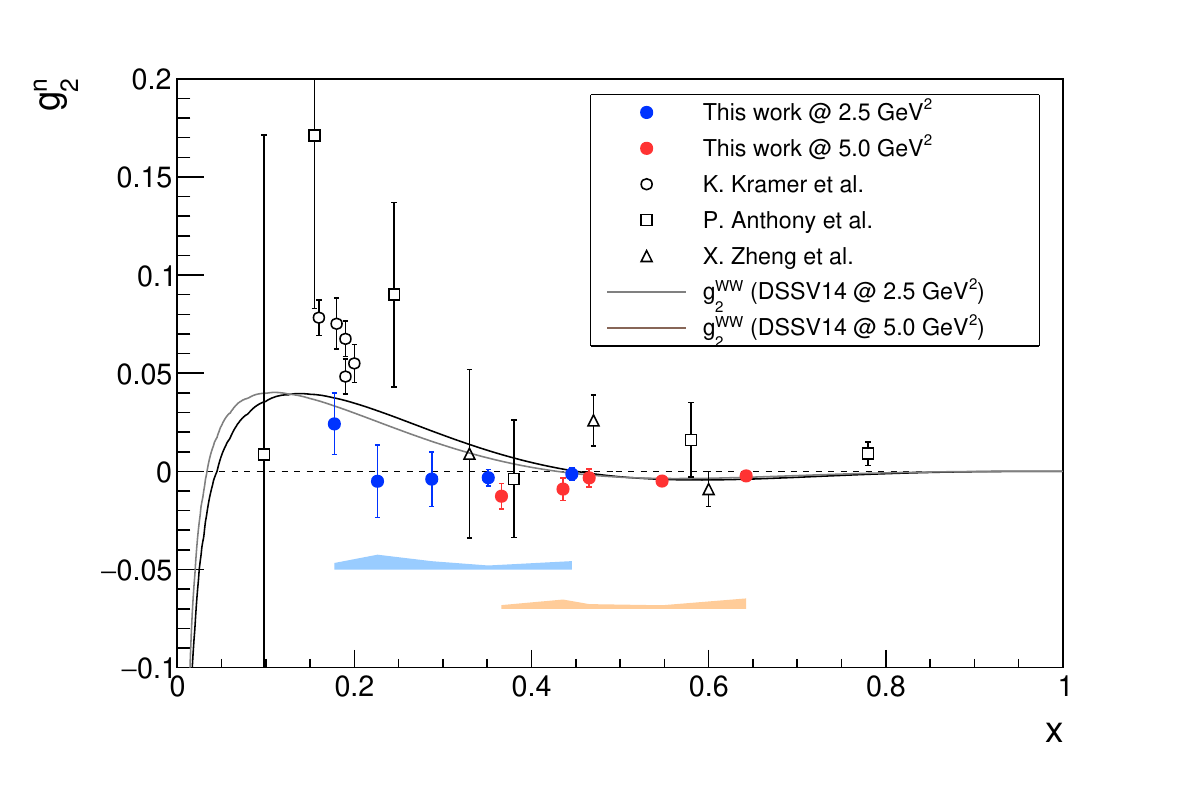}
    \caption{Comparison of the present $g_2^n$ results with previous measurements as a function of $x$. Blue and red filled circles represent the present results at $\langle Q^2\rangle = 2.5$~GeV$^2$ and $5.0$~GeV$^2$, respectively. Error bars indicate statistical uncertainties and the shaded bands below the data points show the systematic uncertainties. The curves show $g_2^{\mathrm{WW}}$ evaluated using DSSV14~\cite{DSSV} at $Q^2 = 2.5$~GeV$^2$ and $5.0$~GeV$^2$.}
    \label{fig:g2n_comp}
\end{figure}

A comparison with the world data is shown in Fig.~\ref{fig:g2n_comp}. The lowest-$x$ point at $\langle Q^2\rangle = 2.5$~GeV$^2$ falls in the region covered by E97-103, which reported a positive deviation from $g_2^{\mathrm{WW}}$ at the lower end of $0.57 \le Q^2 \le 1.34$~GeV$^2$~\cite{Kramer}. At roughly twice that $Q^2$, no positive deviation appears at the same $x$. The present point instead lies below the predictions, though not significantly so within its uncertainty. The tendency becomes more pronounced at larger $x$, where
several adjacent points lie below all four predictions, and the data at $\langle Q^2\rangle = 5.0$~GeV$^2$ show the same behavior. Taken together, our results suggest a negative deviation of $g_2^n$ from $g_2^{\mathrm{WW}}$ over a limited range at lower $x$, at both intermediate and high $Q^2$.

\section{Summary and outlook}
We have presented preliminary results for the neutron spin structure function $g_2^n$ from the E12-06-121 experiment at JLab Hall C, obtained from the unpolarized cross section and the double-spin asymmetries measured in inclusive electron scattering from a polarized $^3$He target. In the DIS region, the data cover $0.18 \lesssim x \lesssim 0.45$ at $\langle Q^2\rangle = 2.5$~GeV$^2$ and $0.37 \lesssim x \lesssim 0.64$ at $\langle Q^2\rangle = 5.0$~GeV$^2$, with $Q^2$ nearly constant within each set. At both values of $Q^2$, $g_2^n$ lies below the $g_2^{\mathrm{WW}}$ prediction over a limited range at lower $x$ and agrees with it at larger $x$, suggesting a negative higher-twist contribution to $g_2^n$ at intermediate and high $Q^2$. The present analysis is restricted to the DIS region. The same data extend into the resonance region at larger $x$. Extending the analysis to that region is the next step toward evaluating $d_2^n$ at the two values of $Q^2$, which will provide a precise benchmark for comparison with lattice QCD.

\end{document}